\documentclass[twocolumn,prl,aps,superscriptaddress,showpacs]{revtex4-1}
\usepackage{mathptmx}
\usepackage{subfigure}
\usepackage{dcolumn}
\usepackage{amsmath,amssymb}
\usepackage{bm}
\usepackage{color}
\usepackage{latexsym}
\usepackage[english]{babel}
\usepackage{times}
\usepackage{psfrag,graphicx}
\usepackage{epsf}
\usepackage{amsfonts}
\usepackage{natbib}
\usepackage{epstopdf}
\usepackage{appendix}
\usepackage[colorlinks=true,citecolor=blue,linkcolor=blue]{hyperref}
\usepackage{ulem}

\begin{document}

\title{Quantum simulation of an extra dimension}

\author{O. Boada}
\affiliation{Dept. d'Estructura i Constituents de la Mat\`eria,
Universitat de Barcelona, 647 Diagonal, 08028 Barcelona, Spain}

\author{A. Celi}
\affiliation{ICFO - The Institute of Photonic Sciences Av. Carl Friedrich Gauss, num. 3, E-08860 Castelldefels (Barcelona), Spain}

\author{J. I. Latorre}
\affiliation{Dept. d'Estructura i Constituents de la Mat\`eria,
Universitat de Barcelona, 647 Diagonal, 08028 Barcelona, Spain}

\author{ M. Lewenstein}
\affiliation{ICFO - The Institute of Photonic Sciences Av. Carl Friedrich Gauss, num. 3, E-08860 Castelldefels (Barcelona), Spain}
\affiliation{ICREA-Instituci\'{o} Catalana de Recerca i Estudis Avan\c{c}ats, 08010 Barcelona, Spain }

\date{\today}

\begin{abstract}
We present a general strategy to simulate a $D+1$-dimensional quantum system using a $D$-dimensional one. We analyze in detail a feasible implementation of our scheme using optical lattice technology. The simplest non-trivial realization of a fourth dimension corresponds to the creation of a bivolume geometry. We also propose single- and many-particle experimental signatures to detect the effects of the extra dimension.
\end{abstract}


\maketitle

{\it Introduction}. 
There is long-standing interest in Physics for the possible existence and effects of extra dimensions. 
This interest was brought by the seminal papers by Kaluza and Klein \cite{kk26} aimed at unifying interactions using the presence of a
fourth spacial dimension. Later on, extra dimensions became a {\sl sine qua non} element for the construction
of string theory \cite{gsw87}. Separately, enormous progress has been made in recent
years to achieve real quantum simulations, that is, to
simulate quantum mechanical models using other well controlled quantum systems \cite{Feynman82}. 
It is now reasonable to investigate to what extent present technology can be used to faithfully simulate a quantum theory living in extra dimensions.

Let us briefly recall recent progress on quantum simulation of condensed matter models using cold atoms \cite{lewe07,bloch08}. By confining atoms to an optical lattice, the Hubbard model may be realized \cite{JBCGZ98}, and the superfluid-to-Mott-insulator transition observed \cite{Greiner02}. Furthermore, several schemes to couple neutral cold atoms to artificial Abelian \cite{JZ03} and non-Abelian \cite{OBSZL05,RJOF05} magnetic and electric fields have been put forth \cite{YCPPPS09,YCPPPS11,SOC,Aidelsburger11}. This opens the door to creating strongly correlated quantum-Hall states with cold atoms \cite{Gemelke10}. 
Although cold atoms are non-relativistic, it is possible to simulate relativistic effects by looking at the low-energy behaviour of some special lattice models --lattice models where the band structure presents Dirac cones, e.g. honeycomb lattices \cite{Zhu2007,Sengstock2010} or lattices dressed with internal degrees of freedom \cite{Goldman09}. Hence, it is not far-fetched that that cold atoms may provide some insight into particle physics models that are not completely understood, such as quantum chromodynamics (see for instance \cite{Montvay97}) or, as presented here, in the analysis of extra dimensions.

{\it General strategy}. The basic idea to achieve a quantum simulation of an extra dimension
consists in engineering the connectivity of the system partly on real dimensions, and partly
on the use of different species for the degrees of freedom (for an alternative approach cf. \cite{Tsomokos10}). 
Let us illustrate this construction
in the case of the simplest quantum mechanical model of a free particle on a hypercubic $D+1$ spacial lattice. 
The Hamiltonian for this system is
\begin{equation}
H=-J\sum_{{\bf q}}\sum_{j=1}^{D+1}  a^{\dagger}_{{\bf q}+{\bf u}_j} a_{\bf q}+ H.c.\, ,\label{hamD+1}
\end{equation}
where $a_{\bf q}$ destroys a degree of freedom at site ${\bf q}$, and the ${\bf u}_j$ stand for $D+1$ cartesian
versors that set the actual connectivity of the lattice. 
Now, we write the $D+1$-position index ${\bf q}$ as the combination of a $D$-dimensional position ${\bf r}$ 
and a separate index $\sigma$ in the extra dimension. That is,  the $D+1$-space is decomposed
in hypersurfaces, that we shall call layers, labeled by the $\sigma$ index, ${\bf q}=({\bf r},\sigma)$, and
the Hamiltonian becomes
 \begin{equation}
 H=-J\sum_{{\bf r},\sigma}\bigl(\sum_{j=1}^{D}  a^{(\sigma)\dagger}_{{\bf r}+{\bf u}_j} a^{(\sigma)}_{\bf r} + a^{(\sigma+1)\dagger}_{{\bf r}} a^{(\sigma)}_{\bf r}\bigr) + H.c.\, \label{hamDsp}.
\end{equation} 
 The operator $a^{(\sigma)}$ can be reinterpreted as a Fock operator for the species $\sigma$. 
The total number of species $N$, $\sigma=1,\dots,N$, corresponds to the number of lattice layers in the extra-dimension. 
The crucial requirement to simulate a $D+1$-dimensional model with a set of $N$ species in a $D$-dimensional lattice is 
that each internal state be coupled to only two other states in a sequential way. 

We here propose two ways of realizing the Hamiltonian of Eq. \ref{hamDsp}. Both methods
make use of Raman transitions in optical lattices but differ in the way of constructing internal
degrees of freedom. We shall refer to the two methods as: \\ {\sl i}) State-dependent lattice; {\sl ii}) On-site dressed lattice.

{\it State-dependent lattice}.
  Let us now discuss the realization of our model in an extra dimension using spin-dependent lattices \cite{Mandel03} (alternative state-dependent lattices
could be obtained by trapping long-lived optical states via superlattice techniques \cite{Gerbier10,Sengstock2010}).
The simplest non-trivial step to construct an extra dimension corresponds to a {\it bivolume} geometry, that is,
the quantum system spans over $N=2$, $D=3$, layers connected through an extra dimension. We shall keep this
example in mind, while providing expressions for $N$ generic layers in $D$ dimensions.

It is possible to induce 
a relative phase in the periodic optical potential seen by hyperfine states with different total angular momentum $F$,
by tuning the angle between the linear polarizations of counterpropagating lasers that form the lattice. 
Hence, the respective minima seen by the atoms get separated by the same phase shift. 
For the explicit example of $^{87}Rb$ loaded in a $3D$ lattice, 
the two available values of $F$, $F=1$ and $F=2$, become two different species,
which correspond to the two layers of a bivolume. 
In general, if we have $N$ different minima and the atomic 
states are long-lived, the system will consist of $N$ copies of  $3D$ layers.

\begin{figure}
\begin{center}
\scalebox{0.35}{\includegraphics{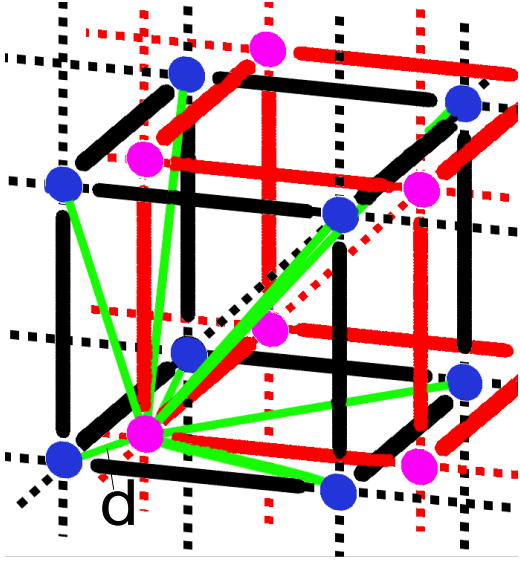}}
\end{center}
\caption{A bivolume geometry is made out of two $3$-dimensional sublattices separated by a displacement vector ${\bf d}$. 
Lattice sites in different colors trap different internal states. Black and red links connect nearest neighbors of 3D sublattices, respectively, and represent the free hopping terms in each sublattice. These tunnelings are due to kinetic energy and do not involve transitions between internal states. Green links connect nearest neighbor sites from different sublattices and are induced by laser assisted Raman transitions. The strength of this coupling depends very strongly on the distance between the pairs of sites.}  
\label{2coupledlat}
\end{figure}

At this stage, there is no hopping between different layers. 
The additional ingredient required to map uncoupled $3D$ hopping 
Hamiltonians to a single $4D$ Hamiltonian is a coupling between different minima or layers. 
The main difficulty is to ensure that each lattice site have the appropriate number of nearest-neighbours. 
In a regular optical lattice this is always the case as Wannier functions --single-particle 
functions living on each lattice site-- at different positions are orthogonal. 
To analyze this requirement, 
let us order arbitrarily the hyperfine states and label them with the index $\sigma=1\dots,N$. We now consider a Lambda configuration between the state $\sigma$ and the state $\sigma+1$ which induces a Raman transfer between the two species, and hence an assisted hopping term between the two sublattices. The effect of the laser is shown in Fig. \ref{2coupledlat} for $D=3$, where for simplicity only one cubic cell of the two sublattices is pictured and the spectator species are omitted. The two sublattices are separated by a displacement vector ${\bf d}$, defined as the smallest distance between a pair of vertices of the two cubes.  Black and red links represent free hopping terms within the sublattices while green links correspond to Raman transitions. In order to make the picture readable only the stimulated hoppings relative to only one site in the $\sigma+1$ sublattice are displayed. For each of the possible hopping terms, the Raman hopping rate is computed as the overlap integral
\begin{equation}
J^{(j)}_{\sigma\sigma+1}= \tfrac{\Omega_{\sigma\sigma+1}}{2}\int {\rm d}^D \textbf{{\bf r}}\, \, \text{w}^*\hspace{-0.6mm}({\bf r}-\boldsymbol{l}^j ) \text{w}({\bf r}-{\bf d})\,,     
\end{equation}
where the $\boldsymbol{l}^j$, $j=1,\dots,2D$, are the  positions of the blue vertices measured with respect to the front-left-down one, $\boldsymbol{l}^1=0$, and w$({\bf r})$ is the Wannier function centered at ${\bf r}$, which for a hypercubic lattice is the product of the $1$-dimensional Wannier $w$ of each Cartesian direction. In a more complicated scenario, as superlattices, the Wannier functions for the two species may be different.

\begin{figure}
\begin{center}
\scalebox{0.65}{\includegraphics{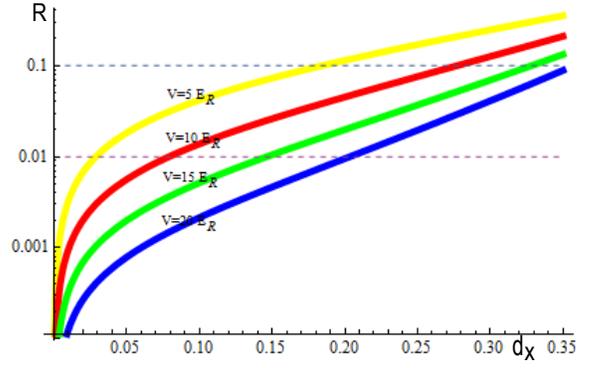}}
\end{center}
\caption{Behavior in log scale of the hopping hierarchy under changing of the maximal Cartesian component $d_x$ of the displacement ${\bf d}$ for different values of the lattice potential $V$. From top to bottom, the curves correspond are to $V=5,10,15$, and $20 E_R$. The dash horizontal lines give the suppression by a factor 10 and 100 of the next-to-leading order coupling. Small values of $R$ allow for the proper bi-volume connectivity of the system.}
\label{hierarchyplot}
\end{figure}

Wannier functions are localized and decay exponentially fast away from each lattice site. A hierarchy between the hopping rate $J^{(1)}_{\sigma\sigma+1}$ and the $J^{(j)}_{\sigma\sigma+1}$, $j>2$, can be easily generated, already for not so small value of $|{\bf d}|/a$, with $a$ the lattice spacing. The behavior of the suppression depends on the depth of the optical potential $V$, and increases with $V$. By using the separability of the Wannier functions, the maximal ratio between the unwanted links and the 
first link can be computed as 1d  problem
\begin{equation}
R\equiv \tfrac{\text{Max}\{J^{(j)}_{\sigma\sigma+1}, j>2\}}{J^{(1)}_{\sigma\sigma+1}} =  \frac{\int {\rm d}x\, w^*\hspace{-0.6mm}(x) w(x+a-d_x)}{\int {\rm d}x  \,w^*\hspace{-0.8mm}(x) w(x-d_x)}\,, \label{ratio} 
\end{equation}
where $d_x$ is the largest Cartesian component of displacement ${\bf d}$. The  optimal scenario at fixed $|{\bf d}|$ is for ${\bf d}$ along the diagonal of the hypercube, i.e. $d_x=\tfrac{|{\bf d}|}{\sqrt D}$. The ratio $R$  is plotted in log scale for typical values of $V$ about ten times the recoil energy, $E_R\equiv \tfrac{(\pi \hbar)^2}{(2 m a^2)}$, in Fig. \ref{hierarchyplot}.  The suppression is very efficient: for $V=20 E_R$ and $d_x= \tfrac a5$ the other hoppings are less $1\%$ of $J^{(1)}_{\sigma\sigma+1}$.  
Under this condition, the effective Hamiltonian is
\begin{equation}
H=-\sum_{{\bf r},\sigma}\bigl(\sum_{j=1}^{D} J a^{(\sigma)\dagger}_{{\bf r}+{\bf u}_j} a^{(\sigma)}_{\bf r} + J' a^{(\sigma+1)\dagger}_{{\bf r}} a^{(\sigma)}_{\bf r}\bigr) + H.c.\, \label{hamstatedep},
\end{equation} 
where for simplicity we assume a uniform hopping between species $J'=J^{(1)}_{\sigma\sigma+1}$,  $\forall \sigma$. The above Hamiltonian coincides with the one of Eq. \ref{hamDsp}, and hence is equivalent to the free hopping Hamiltonian Eq. \ref{hamD+1} in $D+1$-dimensions.

A few considerations are in order. First, we may choose periodic or open boundary conditions for the hopping term in the extra-dimension, by either including or not a Raman stimulated transfer between the $N$ and $1$ states. In the former, our model is equivalent to $D+1$-model compactified on a circle (cf. \cite{kk26,gsw87,Arkani01}). Furthermore, as non-zero constant or ${\bf r}$-dependent phase can be chosen for the hopping rate $J'$, the above set-up allows the simulation of compatifications in presence of non trivial back-ground magnetic fluxes piercing the circle.   
Second, if the interactions are negligible and open boundary conditions are chosen, the only limitation to the ``tickness'' of the extra-dimension is due to the number of hyperfine states available, and to the technical difficulty of coupling them selectively. Indeed, for any value of potential $V$ we can find a displacement ${\bf d}$ sufficiently small such that $R\ll 1$ and Eq. \ref{hamstatedep} holds.  

Once quartic interactions due to binary scattering of atoms are included, the non-zero overlap of the Wanniers of different $D$-dimensional sublattices potentially induces nearest neighbours interactions in the $D+1$-dimension. Again, the separability of the hypercubic lattice allows to express the ratio between the  nearest neighbours and onsite interactions as 
\begin{equation}
(R_U)_{\sigma\sigma+1}\equiv\frac{\alpha_{\sigma\sigma+1}}{\alpha_{\sigma\sigma}} \prod_{j=1}^D \frac{\int {\rm d}x_j\,\left|w(x_j) w(x_j-d_j)\right|^2 }{\int \rm {dx}_j\,\left|w(x_j)\right|^4}\,,
\end{equation}
where $d_j$, $j=1,\dots,D$, are the Cartesian components of the displacement ${\bf d}$, and $\alpha_{\sigma\sigma+1}$ and $\alpha_{\sigma\sigma}$ are the scattering lengths of $\sigma$-$\sigma+1$, and $\sigma$-$\sigma$ collisions, respectively. In most cases, $\alpha_{\sigma\sigma+1}\sim\alpha_{\sigma\sigma}$, hence $(R_U)_{\sigma\sigma+1}=R_U$. In the Table \ref{tableu} the ratio $R_U$ for a displacement along the diagonal of the hypercube, $d_x=d_j$, $\forall d_j$, is given for different values of $V$. As expected, $R_U$ is monotonically decreasing function of both $d_x$ and $V$. By comparison of the numerical result of Table \ref{tableu} with the plot \ref{hierarchyplot}, it turns out that nearest neighbours interactions cannot be disregarded for small value of the optical potential ($V=5 E_R$) while the required hierarchy for both hopping and interaction couplings is realized for a range of displacements $d_x$ at larger values of $V$.  
In fact, a non-trivial $R_U$ may lead to interesting new phenomena like supersolidity \cite{LC70}.
 
\begin{table}[hb]
\begin{center}
\scalebox{.7}{\includegraphics{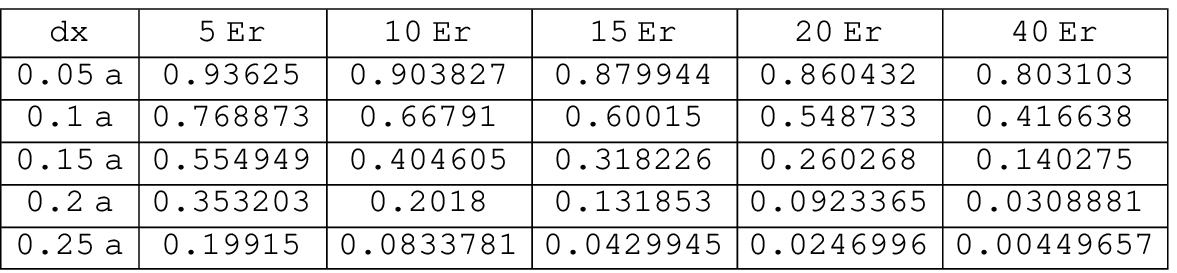}}
\end{center}
\caption{The ratio $R_U$ between the inter-layer and the onsite interaction terms for $D=3$,  as a function of the optical potential $V$ and of the $x$-component of the displacement $d_x=|{\bf d}|/\sqrt 3$. As confirmed by the numerical results, $R_U$ is monotonically decreasing function of $V$ and $d_x$. Only for $V=5 E_R$, $R_U$ is never below 1/10 in the range of $d_x$ compatible with a bivolume hopping term, see figure \ref{hierarchyplot}. The inequality $R_U<1/10$ for $d_x/\sqrt 3\ge a/4$ and $V$ around $10 E_R$ and higher, ensures that a four 3-dimensional layer with negligible inter-layer interaction can be consistently achieved for such values of the potential. For $V$ greater that $20 E_R$ a five 3-dimensional layer is possible.}\label{tableu}
\end{table}
 
 {\it On-site dressed lattice}. In another way to realize of Eq. \ref{hamDsp}, the internal atomic degrees of freedom can be obtained considering hyperfine states with same total angular momentum $F$ and different third component $m_F$. In our case, we are interested in atoms with sufficiently large values of $F$, in order to have several layers in the extra dimension. Very recently, earth-alkali atoms with such property have attracted  a lot of attention from a theoretical \cite{Gorshkov10,Hermele11} and an experimental point of view \cite{Fukuhara07,Martinez09,Stellmer09}. As in the ground state they have total orbital momentum $J=0$, and the nuclear spin $I$ practically does not couple to the dynamics, it follows that $F=I$, and their interaction are SU$(N)$ invariant with $N$ up to 10 for $^{87}Sr$. Such atoms are fermions. Smaller symmetry groups can be achieved also for alkali atoms  \cite{Wu03}. The mixing interaction between species can be realized by optical means as in \cite{Mazza11}.  It is worth noticing that there is no additional difficulty in: {\sl i}) realizing periodic boundary conditions by identifying the species $N+1$ and 1, as this amounts to coupling species 1 and $N$ as well; {\sl ii}) engineering hoppings  with non-trivial phases equivalent to magnetic fluxes piercing the compactified circle.   If 2-body interactions between species are present, the corresponding extended Hubbard model has a very rich phase diagram that is at present under study \cite{inpreparation}.     
 Such scenarios are technically very challenging and involved but potentially feasible using state of the art techniques.
 
{\it Single-particle observables.} We now turn to discuss observables that
can reveal the properties of an extra dimension, whatever experimental technique is used to implement it.
The most direct way  is to consider single-particle effects. The effective dimensionality of a system is revealed by the scaling behavior of observables.
For instance, the two-point correlator free bosonic relativistic field theory in $D$ spacial dimensions, decays 
as $e^{-m|x|}/{|x|}^{(D-1)}$, where $|x|$ is a space-time distance, $m$
is the mass of the field. Such dimensionality dependence can be interpreted as the effect of monogamy of entanglement \cite{mon}.
The more neighbors exist, the more distributed the correlations must be.
In this simple framework, the effect of an extra dimension which can be adiabatically
coupled in the system must translate into an interpolation between the decay exponents in the propagator.
This is managed by a tower of the so-called Kaluza-Klein modes that
bring a series of massive states into the spectrum. We show here that the local {\it density of states}, 
$\rho(E)$, routinely measured in cold atoms experiments \cite{lewe07,bloch08}, displays an analogue behavior once computed for our optical lattice scheme.

To compute $\rho(E)$, we consider the Fourier transform of the free Hamiltonian Eq. \ref{hamDsp} with periodic boundary conditions 
\begin{equation*}
H=- \sum_{\sigma,\sigma'=1}^N\int \tfrac{{\rm d}^D{\bf k}}{(2\pi)^D}\, 
\bigl( 2 J \sum_{j=1}^D \cos(k_j) \delta^{\sigma,\sigma'} + J'  C^{\sigma\sigma'}\bigr)
  a^{\dagger(\sigma)}_{\bf k} a^{(\sigma')}_{\bf k} ,
\end{equation*}
where $C^{\sigma\sigma'}=\delta^{\sigma,(\sigma+1)}+\delta^{\sigma,(\sigma+1)}$ 
(the $N+1$ layer is identified with the first one due to the periodic boundary conditions) is the
the matrix that generates the cyclic group $Z_N$. This Hamiltonian can be diagonalized
in the space of layers,
$H=\sum_{\sigma=1,N} \int \tfrac{{\rm d}^D{\bf k}}{(2\pi)^D}\, E^{D,N}({\bf k},\sigma)  a^{\dagger(\sigma)}_{\bf k} a^{(\sigma)}_{\bf k}$
with
\begin{equation*}
E^{D,N}({\bf k},\sigma)= -2 J \sum_{j=1,D} \cos{k_j} - 2 J' \cos\bigl(\tfrac{2\pi(\sigma-1)}{N}\bigr),
\end{equation*}
where the tower of Kaluza-Klein modes modify the dispersion relation introducing terms proportional to 
the coupling between layers $J'$. It follows that the density of states is
\begin{equation*}
\rho_{D,N}(E)=\tfrac{1}{N}\sum_{\sigma=1,N}\int \tfrac{{\rm d}^D{\bf k}}{(2\pi)^D}\, \delta(E- E^{D,N}_{\bf k},\sigma).
\end{equation*}
First, we observe that for $N\gg1$, the sum can be approximated by the integral over $k_{D+1}\equiv 2\pi(\sigma-1)/N$, and $\rho_{D,N}(E)\rightarrow\rho_{D+1,1}(E)$.
Second, $\rho_{D,N}(E)$ can be computed analytically in the low-energy limit, $E\sim E_{min}=-2(DJ +J')$, for $J=J'$, or for $J'\ll J$, by taking the continuous limits $\cos k_j\to (1-\tfrac 12 k_j^2)$, {\footnotesize $j=1,\dots,D+1$}, or {\footnotesize $j=1,\dots,D$}, respectively. In the latter case, the density of states for $N=2$, in terms of  the one of two uncoupled layers i.e. $\rho_{D,N=1}\propto \tfrac  1J (\tfrac {E-E_{min}}J)^{D/2-1}$, is
\begin{multline*}
\rho_{D,N=2}\sim \tfrac 12 \rho_{D,1}\bigl((1+\tfrac{J'}{E-E_{min}})^{\tfrac D2-1}+(1-\tfrac{J'}{E-E_{min}})^{\tfrac D2-1}\bigr)\\= \rho_{D,1}(1+ \tfrac{(\tfrac D2-1)(\tfrac D2-2)}2  (\tfrac{J'}{E-E_{min}})^2+\dots). 
\end{multline*}    
In particular, the above expression provides an interesting experimental signature for a bivolume, $D=3$.

{\it A many body observable: MI-SF transition.}
In the presence of interactions, the effective dimensionality of the system can be experimentally detected by measuring the location of the Mott-insulator-to-superfluid phase transition. For simplicity, we focus on the bosonic case. It is well known that the ground-state of Bose-Hubbard Hamiltonian in any dimensions presents two phases: for $J/U \ll (J/U)_c$, the ground-state is a Mott-insulator with definite local occupation. For $J/U \gg (J/U)_c$ the ground state is a superfluid state with all the atoms condensed in the single-particle state with null pseudo-momentum. The two phases, Mott-insulator and superfluid, are characterized by the local order parameter $\langle a_{\bf q} \rangle$. As the Mott-insulator ground-state has a well-defined local occupation we have $\langle a_{\bf q} \rangle=0$ everywhere, i.e. no atom number fluctuations. For the superfluid state $\langle a_{\bf q} \rangle \ne 0$, as the occupation is local in momentum and not in position. The critical point $(J/U)_c$ separates the phases with zero and non-zero order parameter, or equivalently, where the symmetry $a_{\bf q}  \rightarrow e^{i\lambda} a_{\bf q} $ is broken and where it is not. This critical point $(J/U)_c$ depends on the dimensionality of the lattice. 

To be more precise, the effective Hamiltonian of a multi-state optical lattice in $D$ spatial dimensions corresponds to
\begin{multline}
 H=-\sum_{{\bf r},\sigma}\bigl(\sum_{j=1}^{D} J a^{(\sigma)\dagger}_{{\bf r}+{\bf u}_j} a^{(\sigma)}_{\bf r} + J' a^{(\sigma+1)\dagger}_{{\bf r}} a^{(\sigma)}_{\bf r}\bigr) + H.c. + \\ + \sum_{{\bf r},\sigma} \tfrac U2 \hat n_{\bf r}^{(\sigma)} \bigl((\hat n_{\bf r}^{(\sigma)}-1) + R_U (\hat n_{\bf r}^{(\sigma+1)}-1)- 2 \tilde \mu\bigr)
 \, \label{hamcompl},
\end{multline}  
where $U=\frac{4 \pi \hbar^2 \alpha}m \int {\rm d}^D {\bf r}\,\left|w({\bf r})\right|^4$, $\hat n_{\bf r}^{(\sigma)} = a^{(\sigma)\dagger}_{\bf r} a^{(\sigma)}_{\bf r}$ is the number operator,  and $\tilde \mu= \tfrac{\mu}U$ is the chemical potential measured in units of $U$. If $R_U\ll 1$ the above Hamiltonian is the Bose-Hubbard Hamiltonian in $D+1$ dimensions. The value of $R_U$ is controlled by the the depth of the optical potential $V$, and the inequality is always satisfied for large enough $V$. 
Thus, we can appropriately tune the scattering length $\alpha$ such that MI-SF occurs for a very large value of $V$.

\begin{figure}
\begin{center}
\scalebox{0.7}{\includegraphics{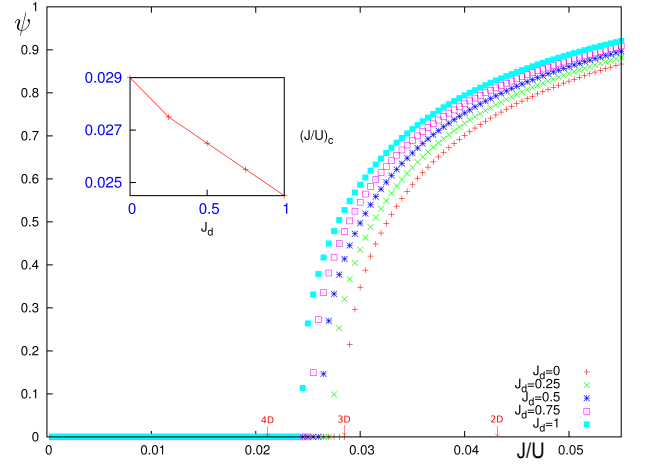}}
\end{center}
\caption{Ground-state order parameter $\psi$ as a function of $J/U$ for different values of coupling $J'$ between the two volumes. For $J'=0$ we obtain the same result as for $D=3$. As the coupling between the layers increases, the critical point $(J/U)_c$ approaches its $D=4$ value for $J'=2J$, which corresponds to $J'=J$, when periodic boundary conditions are chosen. The inset shows the critical point $(J/U)_c$ as a function of the interlayer coupling.}
\label{figMottsf}
\end{figure}

In Fig. \ref{figMottsf} we plot the order parameter computed using a Gutzwiller ansatz as a function of $J/U$ for a lattice made of two $D=3$ layers. Different curves correspond to different values of the coupling between layers, $J'$. As $J' \rightarrow 2 J$, which corresponds to $J' \rightarrow J$ in the case of periodic boundary condition,  the critical value approaches the known value for four dimensions, in mean-field theory. The inset shows the theoretical prediction of the shift in $(J/U)_c$ as the coupling between the two layers increases.

Within the mean-field approximation, the value of $(J/U)_c$ can be computed analytically extending the second order perturbative approach of \cite{Fisher89,VanOsten01} to the Hamiltonian in Eq. \ref{hamcompl} for $N$ layers periodically identified and $R_U=0$. Indeed, due to the periodic boundary conditions in the compact extra-dimension -the lattice is assumed to be sufficiently extended in the other $D$ dimensions such that boundary conditions do not matter- the order parameter is constant (and it can be taken real) on the lattice $\psi\equiv \langle a^{(\sigma)}_{\bf r}\rangle$, $\forall {\bf r},\sigma$. The critical value is found when the symmetric phase $\psi=0$ becomes unstable, i.e. when the $\tfrac {\partial^2 E(J/U,J'/U,\tilde \mu)}{{\partial E}^2}=0$. Such quantity can be computed exactly, within the mean-field approximation, by treating the hopping term as a perturbation at second order. The result is 
\begin{equation}
\alpha \bar U_c= 2 \bar n  +1 \sqrt{(2 \bar n  +1)^2 -1},\label{uc}
\end{equation}
where $\alpha= D/(D+J'/J)$, $\bar U_c=U/(2 D J)$ is the critical value made independent of the connectivity of the hypercubic lattice, and $\bar n$ is the occupation of the Mott state in each species. For $J'=0$, it reduces to the known expression.  Let us stress that Eq. \ref{uc} is not affected by the number of species, $N$. In fact, for $J'=J$ it follows that $(J/U)_c$ for $D$-dimensional model with $N$ layers {\it coincides} with critical value for the $D+1$-Bose-Hubbard. This the case because in the mean-field approximation only local properties like the number of neighbors count, and nothing can be said about global properties like topology and size of one spacetime direction.  Although this prediction cannot be exact --$m$-point correlation functions for $m \ge N$ certainly distinguish whether a spacetime direction is compactified on a circle or not-- we expect that the mean-field result is qualitatively correct, and that a small number $3D$ layers are sufficient to reproduce the $4D$ critical behavior.      

{\it Conclusions}. We have presented a strategy to produce a quantum simulation of an extra dimension. 
The recipes we have proposed to engineer $4D$ models pave the way to the study of novel phases, not accessible in 3D condensed matter world. Furthermore, together with the capacity to simulate the propagation of pseudoparticles in nontrivial gauge fields and in a gravitational background \cite{Boada10}, the present proposal is a step to a complete tool-box for simulating quantum field theory scenarios, in and beyond the Standard Model of particle physics.

{\it Acknowledgements.} We acknowledge the financial support from the Spanish MEC project
TOQATA (FIS2008-00784), 
QOIT (Consolider Ingenio 2010), ERC Advanced Grant
QUAGATUA, EU STREP NAMEQUAM, and from the Alexander von
Humboldt Foundation.


\begin{thebibliography}{10}

\bibitem{kk26}
T. Kaluza, Sitzungsber. Preuss. Akad. Wiss. {\bf 1921} 966 (1921);
O. Klein, Z. Phys A {\bf 37} 895 (1926).

\bibitem{gsw87} M. Green, J. H. Schwarz, and E. Witten, {\em  Superstring theory} (Cambridge University Press. Vol. 2, 1987). 

\bibitem{Feynman82} R.P. Feynman, Int. J. Theor. Phys. {\bf 21}, 467 (1982).  

\bibitem{lewe07} M. Lewenstein {\it et al.}, Adv. Phys. {\bf 56},   243 (2007).

\bibitem{bloch08} I. Bloch {\it et al.}, Rev. Mod. Phys. {\bf 80}, 885 (2008).

\bibitem{JBCGZ98} D. Jaksch {\it et al.}, Phys. Rev. Lett. {\bf 81}, 3108 (1998).

\bibitem{Greiner02} M. Greiner {\it et al.}, Nature, {\bf 415}, 39 (2002).

\bibitem{JZ03} D. Jaksch and P. Zoller, New J. Phys. {\bf 5}, 56 (2003).

\bibitem{OBSZL05}K.~Osterloh {\it et al}, Phys. Rev. Lett. {\bf 95}, 010403 (2005).

\bibitem{RJOF05}J.~Ruseckas {\it et al}, Phys. Rev. Lett. {\bf 95}, 010404 (2005).

\bibitem{YCPPPS09} Y.-J. Lin {\it et al.}, Phys. Rev. Lett. {\bf 102}, 130401 (2009).

\bibitem{YCPPPS11} Y.-J. Lin {\it et al.}, Nat. Phys. {\bf 7}, 531 (2011).

\bibitem{SOC}Y.-J. Lin {\it et al.}, Nature, {\bf 471}, 83 (2011).

\bibitem{Aidelsburger11} M. Aidelsburger {\it et al.},  arXiv:1110.5314 (2011).

\bibitem{Gemelke10} N. Gemelke  {\it et al.},  arXiv:1007.2677 (2010).

\bibitem{Zhu2007}
S.-L. Zhu  {\it et al.},
\newblock Phys. Rev. Lett. {\bf 98}, 260402 (2007).

\bibitem{Sengstock2010}
P.~{Soltan-Panahi} {\it et al.}, Nat. Phys. {\bf 7}, 434 (2011).

\bibitem{Goldman09} N. Goldman {\it et al.}, 
 Phys. Rev. Lett. {\bf 103}, 035301 (2009).

\bibitem{Montvay97}
I.~Montvay and G.~M{\"u}nster,
\newblock {\em Quantum Fields on a Lattice} (Cambridge University Press,
  Cambridge, 1997).
  
\bibitem{Tsomokos10} D. I. Tsomokos, S. Ashhab, and F. Nori,   Phys. Rev. A {\bf 82}, 052311 (2010).

 \bibitem{Mandel03} O. Mandel {\it et al.},  Phys. Rev. Lett. {\bf 91}, 010407 (2003).

\bibitem{Gerbier10} F. Gerbier and J. Dalibard, New J. Phys. {\bf 12}, 033007 (2010).
 
\bibitem{Arkani01}
N. Arkani-Hamed, A. G. Cohen, and H. Georgi, Phys. Rev. Lett. {\bf 86} 4757 (2001).

\bibitem{LC70} A. J. Leggett, Phys. Rev. Lett. {\bf 25} 1543 (1970); G.V. Chester, Phys. Rev. A {\bf 2} 256 (1970). 
  
  
\bibitem{Gorshkov10} A. V. Gorshkov {\it et al.}, Nat. Phys. {\bf 6}, 289 (2010).
 
\bibitem{Hermele11} M.Hermele and V. Gurarie, arXiv:1108.3862. 


\bibitem{Fukuhara07} T. Fukuhara {\it et al.}, Phys. Rev. Lett. {\bf 98}, 030401 (2007).

\bibitem{Martinez09} Y. N. Martinez de Escobar {\it et al.}, Phys. Rev. Lett. {\bf 103}, 200402 (2009). 

\bibitem{Stellmer09} S. Stellmer {\it et al.}, Phys. Rev. Lett. {\bf 103}, 200401 (2009). 

\bibitem{Wu03} C. Wu, J. Hu, and S. Zhang, Phys. Rev. Lett. {\bf 91}, 186402 (2003).

\bibitem{Mazza11} L. Mazza {\it et al.}, \textrm{arXiv:1105.0932} (2011).







\bibitem{inpreparation}
 A. Celi  {\it et al.}, in preparation.

\bibitem{mon} 
V. Coffman, J. Kundu, and W. K. Wootters, Phys. Rev. A {\bf 61}, 052306 (2000).



\bibitem{Fisher89}
M. P. A. Fisher {\it et al.},
\newblock Phys. Rev. B {\bf 40}, 1 (1989).

\bibitem{VanOsten01}
D. Van Osten {\it et al.},
\newblock Phys. Rev. A {\bf 63}, 053601 (2001).

\bibitem{Boada10} O. Boada  {\it et al.}, New J. Phys. {\bf 13}, 035002 (2011).



\end{thebibliography}
\end{document}